\documentstyle[preprint,12pt,aps]{revtex}

\begin{document}
\draft
\title{Is it possible to describe short-range NN
correlations in nuclei on the basis of nucleonic degrees
of freedom only ?}

\author{G.G.Ryzhikh}
\address{Institute for Nuclear Research of Russian Academy
of Sciences, 117312 Moscow, Russia}

\date{\today}

\maketitle

\begin{abstract}
Assuming that NN interaction is modified inside a nucleus
by a specific way, phenomenological effective broad core (BC)
NN potential is constructed. It is shown that increase
in the width of repulsive core lets us describe the
Coulomb form factors of few-body nuclei on the basis
of nucleonic degrees of freedom only.
\end{abstract}

\section{INTRODUCTION}

In the last years, microscopic calculations of the lightest nuclei ($A\le 4$)
on the framework of nonrelativistic quantum mechanics
have achieved a high accuracy.
It becomes apparent that the classic approach built around
two-body potential for free nucleons and nucleonic degrees of freedom
only does not explain many important properties of the nuclei.
More sophisticated approaches developing recently have a number
of refinements going beyond a traditional scheme.
They are strongly interrelated if more general theory
(QCD) is consistently used from the beginning.
But in real calculations in nuclear physics
three general types of the refinements can be presented:
\begin{enumerate}
\item Modification of interaction between the nucleons in nuclei
through changing of the pair-wise NN forces themselves and
generation of the multi-nucleon forces.
\item Consideration of non-nucleonic degrees of freedom
(meson, $\Delta$-isobar, quark-gluon etc.).
\item Consistent relativistic description of the process involved.
\end{enumerate}
Just corrections of the first type attract a particular attention because
they can be done in terms of traditional computational scheme.
It is not surprising that the various models
of 3N forces are very popular for solution of the
problems in the theory of few-body systems (for example,
the underbinding of 3- and 4-nucleon systems \cite{3c}).
However, some problems can not be explained on the
framework of pure nucleonic degrees of freedom even if the 3N
forces are taken into account. One of the best known example is
a manifestation of strong short-range NN correlations
in the Coulomb form factors of $^3$H, $^3$He and $^4$He
(see, for example, \cite{3c,SR_34,3b,ff3}).

To explain the short-range correlations in
few-body systems the other types of refinements are
invoked.
Many calculations \cite{3c,SR_34,3b,3mec}  show that
the contribution of two-body meson exchange currents
(MEC) can explain form factors of three- and four-nucleon systems.
However, there are some limitations and open problems
in such explanation.

Strange as it may seem, a weak point in modern nonrelativistic approaches is
the consistent description of deuteron structure. The recent experiments
for the measurements of $T_{20}$ polarization \cite{t20} have allowed
to extract experimentally $C0$ form factor of the deuteron.
It turns out that the theoretical consideration of
the charge current with MEC (where all ingredients of
meson exchanges are the same as in the case of three-body system)
destroys the description of the deuteron charge form factor
obtained in the impulse approximation \cite{SR_2,Henning}.
Thus, the present calculations are not yet able to simultaneously
explain the data for two- and three-body charge form factors
on the equal ground \cite{Henning}.

The next problem is that the high values of cutoff masses $\Lambda$
described meson-nucleon vertexes must be used to explain a
large magnitude of two-body exchange currents in three-body form factors,
that is in contradiction with modern nucleon structure calculations
(see, for example, \cite{Machleidt,vertex,skyrme}).
Besides, very high contribution of
MEC to the charge form factor at moderate momentum transfer seems
to be hardly consistent to the common practice
in nuclear theory (Siegert theorem). It is well known that
MEC presents a relativistic correction to the charge one-body current
by the same order of magnitude as the relativistic correction to
the wave function itself. Therefore, the consistent relativistic
description of NN interaction, few-body wave function and
the process of interest is required if the contribution of
MEC (in particular the isoscalar one) is large.
This conclusion is strongly supported by the recent study of
two-body system in relativistically covariant approaches.
As was shown in \cite{Tjon,Arnold,Arenhovel}, the
relativistic description of two-nucleon system considerably changes
its electromagnetic properties and
reduces substantially the contribution of MEC to the charge form
factor of the deuteron \cite{Tjon}.

It is believed that just the modification of the nucleon
interaction inside the nuclei makes possible to solve the problem
of strong NN correlations more simply.
The small influence of 3N forces to the form factors
of $^3$H and $^3$He in the earlier calculations
(see, for example, \cite{3b}) is probably connected
with dominantly long-ranged structure of these semiphenomenological forces.
It would appear natural that the modification of the short-range part
of interaction is essential to improve the description of
form factors at moderate and high momentum transfer.
Consistent microscopic calculations \cite{Kolk,Glockle} actually show
that the main long-ranged components of 3N forces are canceled and,
along with it, the shorter ranged forces arise (which,
however, have an uncertain parameter \cite{Kolk}).

This work consists in an attempt to explain the charge form factors of
the lightest nuclei by the phenomenological modification of
NN interaction only.
The general description of the model is given in
Section 2. The results and discussion are presented in
Section 3.

\section{THE MODEL}

As was mentioned above, NN interaction in nuclei is changed
(see also discussion in \cite{Migdal,Brown})
comparatively to the "vacuum" case.

The aim of this work is to study the
possibility of description of short-range
NN correlations in the lightest nuclei by
the phenomenological modification of NN potential only.

The main features of this scheme are the following:
\begin{enumerate}
\item  The modification of  pair-wise NN interaction only is considered.
By this means true multi-nucleon forces are beyond the scope of the paper.
This assumption seems to be justified here because the
effects from true multi-nucleon forces are hardly separated in such
integrated characteristics as charge density of a nucleus.
In addition a lot of new parameters in the general case results
in great uncertainty in phenomenological analysis.
\item
It is suggested that the interaction is modified inside the nuclei
so that the low-energy NN phase shifts do not
change practically relative to free-nucleon case.
In other words, the common "power" of interaction is kept a little
affected in nuclei in our model.
Indirect evidence that similar situation can be realized in
reality presents microscopic
study of three-body modification of free NN forces.
Thus, in \cite{afnan} was shown
that many components of tree-nucleon forces are
canceled in consistent approach, so that
overall effect of such modification for the
binding energy of $^3$H is very small.
On the other hand, just the modification of "high"-energy part of
two-body interaction leads to the significant
changes of the properties connected with high
momentum transfer that is the most important for our consideration.
\item Since an aim of the paper is the qualitative study of the problem,
the model is restricted to the simplest case of
pure central NN potential. It is known \cite{Af_Tan,he} that
this simplification does not change significantly  the properties
of interest for us, in particular, the charge form factor
$F_{ch}$ and the root mean squared radius $(r^2)^{1/2}_{ch}$,
because the role of non-central components
in the corresponding matrix elements is small (there is no transition
from $S$ to $D$ components of the wave function).
This restriction, however, does not make us possible to
investigate the transversal $M1$ form factors of two- and tree-body
systems because the interference between $S-S$ and $S-D$
transition in the matrix elements is of primarily importance
for these observables \cite{3c,ff3}.
\end{enumerate}

Similar type of simplified (effective) potentials is widely
used in the microscopic calculations \cite{Saito} of nuclear properties
in such approaches as Resonating Group Method (RGM), Generator
Coordinate Method (GCM) , Hartree-Fock model (HF) etc.
These forces are called effective
to stress that some effects from the
nuclear medium are already taken into account by their construction.
Usually, the NN interaction in such an approach is
constructed to fit some properties of few-body systems \cite{Af_Tan,effpot1}
and, may be, nuclear matter too \cite{effpot2}.
However, no one from these potentials allows to describe
quantitatively Coulomb form factors of tree-
and four-nucleon nuclei in the region of its second maxima.
The work \cite{Kolesnikov} had the most success.
In this work was pointed out that to explain the position
of the second maxima of $^3$He and $^4$He form factors it should
be assumed that the NN interaction has a very wide repulsive core.
There are still further reasons for that.
For example, it was shown in \cite{he,our_nuc} that the "hole" in
the nucleon density of $^4$He is much broader than it follows
from the calculations with the realistic NN forces. More
strong repulsion (even with accounting the contemporary 3N forces)
is needed also to explain the density of
nuclear matter in the saturation point \cite{Brown,Machleidt,manybody}.
On the other hand, the theoretical models \cite{sig_om,pion} developed to
study NN interaction in the nuclear matter really show
that the temperature and density plays the role
of additional repulsion.

Thus, we try to construct our effective NN forces by
the increasing of the width of repulsive core in
some "realistic" central potential fitted to the two-body data
(for example, in potential like $S3$ from the work \cite{Af_Tan}).
Since the minor role of the NN interaction in the states
with odd values of partial angular momentum $L$ in all two-body subsystems
for the structure of $S$-shell nuclei, only
potential for the even partial states is constructed.

For the sake of convenience the gaussian parameterization of
NN interaction is used:
\begin{equation}
V(X)=\sum_{i=1}^{N}C_i(X)\exp(-\alpha_i(X) r^2)   \label{gaus}
\end{equation}
Coefficients $C_i$ and $\alpha_i$ for given spin-isospin
configuration $X$ ($X=$$^3\!E$ or $^1\!E$ in our case)
are founded by fitting a set of calculated
observables to the experimental data. Since the three-nucleon
system is the object of main interest for us,
the binding energy $B$, the root mean squared charge radius
$(r^2)^{1/2}_{ch}$ and
the charge form factor $F_{ch}$ for $^3$H is chosen for that.
In addition, following arguments given above,
the low-energy phase shifts for $n$-$p$ scattering and the binding
energy for the deuteron are also incorporated in the fitting procedure.

Form factor of the deuteron and the main properties of $^3$He and
$^4$He are calculated and compared with experimental data
after the construction of effective potential.

In the calculation of wave functions of few-body systems
the variational method based on the multidimensional
gaussian non-orthogonal basis is
applied. This method has been proposed in \cite{Kukulin}
and since that it was widely accepted for the
high-precision calculations of many-body nuclear problems
(see, for example, \cite{our_nuc,Suzuki,Kamimura} and
references therein). The detailed information about the used
scheme can be found in \cite{34bprog,Ergash,Kolesnikov}
for the tree-body systems and in \cite{Kolesnikov,34bprog} for
the four-body one. In the calculation
the interaction in the odd two-body states ($^3P$ and $^1P$) is
considered as equal to zero.

\section{THE RESULTS AND DISCUSSION}

Our calculations are shown that only a priori "knowledge"
of the core width makes possible to construct a desirable potential
by fitting procedure in view of numerical instabilities of
many-parameter minimization.
That is why even indirect information discussed above is of primarily
importance.

The derived values of parameters $C_i$ and $\alpha_i$ are given
in Table I.
Five gaussian parameterization ($N=5$ in eq.~(\ref{gaus}))
with a common set of nonlinear parameters for triplet
and singlet NN potentials is found to be sufficient to meet
all our requirements.
It shoul be noted that fixation even low-energy NN phase shifts
impose so strong restriction on the constructed two-body forces that
the only one real parameter can be
considered as relatively free until the information
about $^3$H charge form factor come into play.
It is the repulsive core width in NN forces can change validly
the two- and many-body charge density at short distances.
Therefore, a big number of parameters
given in Table I is nothing but consequence of chosen parameterization.

The constructed potential is shown in Fig.1a and Fig.1b by the
solid line. One from precise phase-fitted realistic
potential (Nijm.II from \cite{$NN$pot}) in the respective states is
shown by the dashed line.
It is very important that the constructed potential has a repulsive
core ($r_c\simeq 1 fm$) about 1.5--2 times broader than that for
the realistic one ($r_c\simeq 0.5-0.6 fm$).
Thus, we will call it further as Broad Core (BC) potential.

It is interesting to note that the use of potential with
broad core to fit the low-energy NN phase shifts leads
automatically to the appearance of additional repulsive
bump at the moderate distances $r\simeq 2.5 fm$. This is
because the low-energy phase shifts determine
some "mean size" of two-body system and for that it is necessary
to restrict the attractive well from the side of large distances.

In Figs.2 the phase shifts for the BC-potential in the
$^3S_1$ and $^1S_0$ channels are shown.
It turns out that in spite of strong modification of
NN potential needed to fit charge form factor of $^3$H
it is possible to describe quite well NN phase shifts
up to $E_{Lab}\simeq 100\ MeV$.
More strong repulsion at the short distances
leads to that the calculated phase shifts  pass above the
experimental points as the energy increases.
However, the fixation of low-energy part of NN interaction
is found to be sufficient for the rather accurate description
of important static properties of nuclei with $A=2-4$.

In Table II the root mean squared charge radii and the binding energies
of few-body systems are given. For $^3$H and $^3$He the
results of calculations with and without inclusion of $S'$-component
(i.e. component with the mixed permutational symmetry) are shown.
Just the inclusion of this component into the variational basis
leads to the sufficient increasing of the binding energies and
makes possible to explain the experimental difference in
$(r^2)^{1/2}_{ch}$ and $F_{ch}$ (see below) between the two isomeric
three-body nuclei. Such high role
of the mixed symmetry configurations
(with the weight $P\simeq 1.4$\%) is connected with the strong
difference between the NN forces in $^3\!E$ and $^1\!E$
states (especially in the width of repulsive core).

In Figs.3 the charge form factors of $^3$H and $^3$He are
compared with the new experimental data \cite{ff3}.
It is interesting to note that the $S'$ component leads to
the increasing of $^3$He form factor and,
respectively, to the decreasing of $^3$H form factor in the
region of their second maxima. As a result, the quality
of description of both form factors is very high in spite of
that the only $^3$H form factor has been fitted.

The calculation of $^4$He structure was done in the
basis of symmetric component only. Therefore, the accurate binding
energy for BC-potential is somewhat higher than that given
in Table II. At the same time, it is believed that both
charge root mean squared radius and form factor of $^4$H is
described rather accurate in this truncated basis because, in this case,
the number of protons is equal to the number of neutrons.

The Coulomb $C0$ form factors of the deuteron and $^4$He (the
information about which was not used
under the construction of BC-potential)
are shown in  Fig.4a and Fig.4b. The $^4$He form factor
is in a quite good agreement with experimental data \cite{ff4} at the wide
interval of momentum transfer except the range of
second maximum. This fact may be interpreted as
that the role of fourth particle to the modification of NN
interaction is considerably less than the third one (or, in
other words, the relative smallest of $4N$ forces).
The description of the deuteron form factor is good up to
$q^2\simeq 10\ fm^{-2}$. For the higher momentum transfer the
theoretical curve lies above the experiments. This result seems to be
quite reasonable because the effective NN potential
is used in the calculation.

It is very interesting that all Coulomb form factors
for $^2$H,$^3$H,$^3$He and $^4$He calculated in
impulse approximation for BC-potential are in
a perfect agreement with those for realistic NN forces
calculated with inclusion of MEC \cite{SR_2,SR_34}. Thus,
the modification of free NN potential in the presence of
the others nucleons results in the similar
modification of nuclear charge density
as that one produced by the meson exchange currents.
This agreement seems to be not casual because the same
meson exchanges result both in NN forces and in two-body charge density.
That is why the theoretical predictions for the different nuclear forces
(provided the same input information ---  NN phase shifts)
are closed together if considered in a consistent manner
with all meson exchanges \cite{ff3}.

To make the comparison of the calculated results with experimental
data more correct it is important to consider the contribution
of two-body exchange currents in proposed approach too. If two-body
term remains dominant in the second maxima of charge form factors of
few-body systems as before then the agreement with experiment will
break down. Fortunately, this is not the case. Our estimations
show that the absolute magnitude of MEC reduces by the using of
BC potential as compared with the calculations used realistic NN forces.
The reason is that the contribution of MEC is defined as some integral from
the short-ranged Yukawa function \cite{3mec,Riska}:
$Y_1=(1+x)\exp(-x)/x$,
where $x=\mu r$, and $\mu$ is the mass of corresponding meson.
As a result, the magnitude of two-body current drops with
the rising of "hole" in two-body wave function at short distances.
Such effect was shown in the calculation with phenomenological
wave function written in correlated form \cite{3mec,Riska}:
\begin{equation}
U=N_S\prod_{k<l}\left(1-\exp(-\gamma^2r_{kl}^2)\right)^{1/2}
\cdot\exp\left(-\alpha^2/2\sum_{m<n} r_{mn}^2\right)
\end{equation}
In that case the magnitude of MEC at moderate momentum transfer
in three- and four-body charge form factor decreases with the
parameter of two-body correlation $\gamma$.
As to the real calculations with BC-potential, so, for example,
the contribution of MEC to the region of second maximum for charge
form factors of three-body nuclei decreases of about 1.5 times
relatively to that for realistic NN potentials \cite{SR_34,3mec}.
As the one-body part of charge form factor at second
maximum in our approach is strongly increased (see Figs.2)
so the relative importance of two-body current (even calculated
with standard parameterization of meson-nucleon vertexes, see below)
is greatly reduced.

What is more, there is another source of reducing of meson-exchange
contribution to the charge density. As was mentioned above,
modern nucleon structure calculations
(see, for example, \cite{Machleidt,vertex,skyrme}) obtain much lower
meson scales than ordinary used to fit two- and few-body data.
This fact presents a good argument in support of developed model.
On the one hand, just the decrease in cutoff masses $\Lambda$
(in particular $\Lambda_{\pi NN}$)
in one-boson exchange model \cite{Machleidt} leads to the
increasing of repulsive core in NN potential.
On the other hand, soft $\pi NN$ form factor
($\Lambda_{\pi NN}=0.6-0.8$GeV instead of $\Lambda_{\pi NN}\simeq 1.2$GeV
commonly used in the calculations) results in pronounced reduction of
two-body contribution to the electromagnetic
form factors of the lightest nuclei.
Thus, in the proposed approach, magnitude of MEC
can be considered as a correction to the nonrelativistic
one-body charge current up to high momentum transfer.
This situation seems to be much favorable than that one in
the few-body calculations with realistic NN forces
of standard type.

The proposed modification of NN potential is quite similar to that one
obtained in the ref.\cite{sig_om}. In this work the
calculation of NN potential coming from the exchange of
one $\sigma$- or $\omega$-meson, inside nuclear matter
(in $\sigma$-$\omega$ model of Walecka \cite{Walecka}),
taking into account vacuum polarization effects
was performed.
The increasing of the core width at small distances and
generation of Friedel-like oscillations at large ones
was found as a consequence of finite nuclear density.
It should be noted that these
oscillations \cite{Friedel} are rather common phenomena
in the fermion systems.
Recall, that the additional
oscillation in BC-potential at the distance $r\simeq 2.5\ fm$
arises in our approach
as a direct consequence of the broadening of the core to
compensate changing of the low-energy NN phase shifts.
The position and the magnitude of this bump is in agreement
to that found in \cite{sig_om}.
At the same time, it is evident that the details of effective potentials at
higher distances ($r \ge 3-4\ fm$) can not be reconstructed accurately
from the phase shifts fitting. Therefore, the consistent microscopic
description only will be able to derive the accurate behaviour
of effective NN potential at this range.

\section{CONCLUSION}

An attempt to solve some problems of description of strong NN correlations
by the modification of NN potential inside the nuclei is done in the paper.
Our analysis shows that the increasing of the
width of repulsive core makes it possible to describe naturally
the strong short-range correlations in nuclei.
On this ground an effective central broad core NN
potential fitted to the low-energy $S$-wave $n$-$p$ phase shifts is
constructed.
It is shown that this potential makes it possible to
describe quantitatively the charge form factors
of the few-body systems. The attractive feature
of the model is the pronounced reduction both the relative and
absolute contribution of MEC to the charge form factors
of the lightest nuclei.
It is of interest to expand the approach by taken into
account the non-central components of NN forces and to include
into consideration the odd two-body partial states.
It will permit to study the magnetic form factors of few-body systems
and the properties of nuclear matter.

Obtained modification of NN potential is in agreement
with the  qualitative prediction of ref.\cite{sig_om} and with
the small values of cutoff masses in the meson-nucleon
vertex form factors \cite{Machleidt,vertex,skyrme}. At the same
time, the proposed approach is phenomenological. It is of primarily
importance to investigate quantitatively the nucleon-nucleon
interaction in nuclei
on the basis of consistent microscopic approach. It is evident
that the potential derived, for example, from quark-gluon
dynamic must have much more complicated structure. In particular,
strong nonlocalities, energy and momentum dependence, non-central
and many-body terms inevitably arise in the accurate microscopic
consideration. However, there is a reason to hope that the main
features
of NN interaction inside the nuclei (especially the broadening of repulsive
core) are correctly mentioned in the work.

The constructed potential results in the best description
of many aspect of structure of the lightest nuclei amongst the known
effective potentials. It turns out that it can be successfully
used in the microscopic approaches like RGM to accurate
account of the short-range correlations in more "heavy" nuclei
than those are considered in the paper.

       I would like to thank
N.N.Kolesnikov, R.A.Eramzhyan and V.I.Kukulin
for the useful discussions and P.P.Zakharov,
V.I.Tarasov, E.M.Tursunov and V.N.Pomerantsev for the help in the
work and for kindly given opportunity to use the developed computer
codes.

\pagebreak

\begin{table}
\caption{Parameters $C_i$ (in $MeV$) and $\alpha_i$
(in $fm^{-2}$) of BC-potential given in eq.(1) for
$^3\!E$ and $^1\!E$ states.}
\begin{tabular}{l|c|c|c|c|c} \hline
$i$ & 1 & 2 & 3 & 4 & 5 \\ \hline
$C_i(^3\!E)$ &1067 &3530 &--3595 &1475 &--553.6 \\
$C_i(^1\!E)$ &1416 &3309 &--3499 &1175 &--392.4 \\
$\alpha_i$ &5.391 &1.197 &0.7368 &0.3602 &0.2651 \\ \hline
\end{tabular}
\end{table}

\vspace{5mm}

\begin{table}
\caption{Binding energies $B$ (in $MeV$) and root mean squared
charge radii $(r^2)^{1/2}_{ch}$ (in $fm$) of few-body nuclei. $S$ and $S'$
denotes the choice of variational basis in pure symmetric
form ($S$) or as a state with mixed permutational symmetry ($S'$).}
\begin{tabular}{l|c|c|c|c} \hline
 & $^2$H & $^3$H & $^3$He & $^4$He  \\ \hline
$B(S)$ &2.225 &7.5 &6.8 &26.2 \\
$B(S+S')$ & &8.9 &8.2 & \\
$B(exp.)$ &2.225 &8.48 &7.72 &28.3 \\ \hline
$(r^2)^{1/2}_{ch}(S)$ &2.14 &1.85 &1.87 &1.64 \\
$(r^2)^{1/2}_{ch}(S+S')$ & &1.75 &1.89 & \\
$(r^2)^{1/2}_{ch}(exp.)$ &2.10 &1.76 &1.95 &1.67 \\ \hline
\end{tabular}
\end{table}

\vspace*{4cm}

\pagebreak

Figure 1. (a) The NN potential in $^1S_0$ partial
state. The BC-potential is plotted by solid line,
the realistic Nijm.II potential is plotted by dashed line.
(b) The same as in Fig.1(a) for $^3S_1$ partial state.

\vspace{5mm}

Figure 2. Triplet ($^3S_1$) and singlet ($^1S_0$)
$n$-$p$ phase shifts for BC-potential.

\vspace{5mm}

Figure 3. (a) The charge form factor of $^3$H calculated in the
impulse approximation for BC-potential in the pure symmetric basis
(dashed line) and in the complete basis (solid line). (b) The charge
form factor of $^3$He.

\vspace{5mm}

Figure 4. (a) C0 form factor of $^2$H
calculated for BC-potential in the impulse approximation.
(b) The charge form factor of the $^4$He.

\end{document}